\documentclass{ifacconf}

\usepackage{graphicx}      
\usepackage{natbib}        
\usepackage{xcolor}
\usepackage{tikz}
\usepackage{listings}
\usepackage{url}
\usepackage{amsmath, amssymb}
\usepackage{bm}
\usepackage{diagbox}
\usepackage{booktabs}
\usepackage{siunitx}

\usepackage{fancyhdr}
\usepackage{nopageno}

\begin{document}
\begin{frontmatter}

\title{Identifiability and physical interpretability of hybrid, gray-box models - a case study} 

\author[First]{M. Hotvedt} 
\author[First,Second]{B. Grimstad} 
\author[First]{L. Imsland}

\address[First]{Engineering Cybernetics Department, NTNU, Trondheim, Norway (e-mail: \{mathilde.hotvedt, lars.imsland\}@ntnu.no)}
\address[Second]{Solution Seeker (e-mail: bjarne.grimstad@solutionseeker.no)}

\begin{abstract}                
Model identifiability concerns the uniqueness of uncertain model parameters to be estimated from available process data and is often thought of as a prerequisite for the physical interpretability of a model. Nevertheless, model identifiability may be challenging to obtain in practice due to both stochastic and deterministic uncertainties, e.g. low data variability, noisy measurements, erroneous model structure, and stochasticity and locality of the optimization algorithm. For gray-box, hybrid models, model identifiability is rarely obtainable due to a high number of parameters. We illustrate through an industrial case study -- modeling of a production choke valve in a petroleum well -- that physical interpretability may be preserved even for non-identifiable models with adequate parameter regularization in the estimation problem. To this end, in a real industrial scenario, it may be beneficial for the model's predictive performance to develop hybrid over mechanistic models, as the model flexibility is higher. Modeling of six petroleum wells on the asset Edvard Grieg using historical production data show a 35\% reduction in the median prediction error across the wells comparing a hybrid to a mechanistic model. On the other hand, both the predictive performance and physical interpretability of the developed models are influenced by the available data. The findings encourage research into online learning and other hybrid model variants to improve the results. 
\end{abstract}


\end{frontmatter}

\section{Introduction}
\begin{figure}
\centering
\includegraphics[width=6.cm]{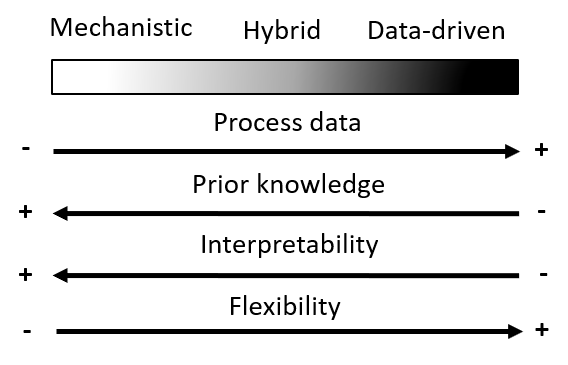}   
\caption{Gray-scale of mathematical models, ranging from mechanistic to data-driven models.} 
\label{fig:gray-scale}
\end{figure}

\thispagestyle{fancy}
\chead{\textit{2021 M. Hotvedt, B. Grimstad, L. Imsland. This work has been accepted to IFAC for publication under a Creative Commons Licence CC-BY-NC-ND}}
\pagenumbering{gobble}
Mathematical modeling of physical processes is an important aspect of many engineering fields and may aid in the analysis and prediction of a process response to changes in state and control variables. Therefore, a model should be a good representation of the underlying process. However, during mathematical modeling, there is often a compromise between complexity and intractable models \citep{Zendehboudi2018}. Usually, a higher model complexity yields a better representation of the process but is harder to solve. Another trade-off is how much prior knowledge should be incorporated into the model and how much should be learned from process data. Typically, mathematical models may be placed on a gray-scale ranging from mechanistic to data-driven models, or, from white- to black-box models respectively, see Fig.~\ref{fig:gray-scale}.

Mechanistic models are built from first-principle equations, with possible empirical closure relations, and require considerable understanding of the physical behavior of the process. These models are often of high complexity and call for simplifications to be computationally feasible. Data-driven models are constructed from generic mathematical equations fitted to process data and require no prior knowledge about the process. These model types have a high degree of \emph{flexibility}, which is the ability to adapt to arbitrarily complex patterns in data. Therefore, contrary to mechanistic models, data-driven models may capture unmodeled or unknown process behavior as long as these are reflected in the available data. However, data-driven models are data-hungry and require the data to be sufficiently rich to represent the process behavior appropriately. If care is not taken, overfitting is a frequent outcome resulting in poor extrapolation power to future process conditions \citep{Solle2016}. As a result of using first-principle equations, mechanistic models are typically better at extrapolation. 

Hybrid modeling combines mechanistic and data-driven modeling techniques and attempts to preserve the advantages while diminishing the disadvantages of the two approaches. There are several ways to construct hybrid models and much literature on the topic (e.g. \citet{Psichogios1992, Kramer1992, Xu2011, Zendehboudi2018, Bikmukhametov2020b, Hotvedt2020}). Hybrid modeling may be approached from either side of the gray-scale in Fig.~\ref{fig:gray-scale}. Models on the white side of the gray-scale typically have a higher degree of \emph{interpretability} than models on the black side. In literature, there are numerous definitions of, and several aspects connected to interpretability. According to \citet{Roscher2020} is interpretability the ability to present a model in understandable terms to a human. The focus in this article is on the aspect \emph{physical} interpretability. A model will be physically interpretable if 1) its output behavior is in line with existing physical principles, and 2) any model parameter representing a physical quantity has a physically feasible value. From a process analysis perspective, physically interpretable models are of great value as the models are often better at extrapolation and the analysis of the model response to changes in variables is simplified. This promotes trust in the model. To this end, this article will focus on hybrid models on the white side of the gray-scale in Fig.~\ref{fig:gray-scale}. A mechanistic model is used as a baseline and a data-driven model is inserted to increase model flexibility. 

According to \citet{Deconinck2017}, a prerequisite to obtaining physically interpretable models is model \emph{identifiability}. With this property, uncertain model parameters may be \emph{uniquely} determined from the available process data \citep{Goodfellow2016}. However, when a model is moved on the gray-scale from mechanistic towards data-driven, the number of parameters increases and the model generally becomes non-identifiable given the available data. This is particularly the case for deep neural networks which are often designed to have a complexity above the interpolation threshold. Furthermore, identifiability is influenced by both stochastic uncertainties, e.g. noisy data and stochasticity of the estimation problem, and deterministic uncertainties, e.g. erroneous model structure and local optimization algorithms, making most models non-identifiable in practice. For these reasons, in the data-driven modeling domain, a common practice is to seek a model that generalizes well and has a high degree of flexibility, rather than seeking identifiable models. On the other hand, data-driven models typically have non-physical parameters. Ensuring unique parameter values are therefore not as important as for hybrid and mechanistic models where there exist physical parameters whose values should be physically consistent to obtain physically interpretable models. Identifiability and interpretability have been topic for many studies, e.g. \citep{Brun2001, Raue2009, Deconinck2017, Brastein2019}, but few have studied this for hybrid, gray-box models.  

This article will study the identifiability, physical interpretability, and predictive performance of a mechanistic and a hybrid petroleum well production choke valve model. The first two concepts are investigated through a synthetic case study where we have perfect information about the underlying process. Thereafter, the two latter concepts are examined in an industrial scenario where real and historical production data from six petroleum wells on Edvard Grieg \citep{EG} are utilized in the model development. 

\section{Parameter estimation problem}
For all models on the gray-scale, parameter estimation is essential to obtain adequate model predictions of the process. Consider a dataset $\mathcal{D} = \{\bm{x}_i, y_i\}_{i=1}^n$ with $n$ observations of the process state and control variables $\bm{x}_i\in\mathbb{R}^d$, also called explanatory variables, and target variable $y_i \in \mathbb{R}$. Assume the process to be described by 
\begin{equation}\label{eq:model}
    y_i = f(\bm{x}_i; \bm{\phi}) + \epsilon_i, \quad \epsilon_i \sim \mathcal{N}(0,\sigma_{\epsilon}^2),
\end{equation}
where $\hat{y}_i = f(\bm{x}_i; \bm{\phi})$ is a general steady-state model with parameters $\bm{\phi}\in\mathbb{R}^m$, and $\epsilon_i$ is normally distributed measurement noise. In general, the parameter estimation problem may be posed as an optimization problem
\begin{equation}\label{eq:optProb}
    \bm{\phi}^* = \arg \min_{\bm{\phi}} J(\bm{y}, \hat{\bm{y}}),
\end{equation}
where $J$ is a scalar, non-negative objective function, and $\bm{y}, \bm{\hat{y}} \in\mathbb{R}^n$ are the observations of target variable measurements and estimates respectively. A common approach to solving the estimation problem is through least squares estimation, which is equivalent to maximum likelihood estimation (MLE) of the model in \eqref{eq:model}. Denoting the design matrix by $X \in \mathbb{R}^{n\times d}$ with rows $X_{i, *} = \bm{x}_i$, the MLE may be written compactly as
\begin{equation}\label{eq:MLE}
    \bm{\phi}^*_{MLE} = \arg \min_{\bm{\phi}} \left(\bm{y} - f(X; \bm{\phi})\right)^{\top}\left(\bm{y} - f(X; \bm{\phi})\right).
\end{equation}
This is a nonlinear, nonconvex optimization problem that generally has several solutions, which all fulfill
\begin{equation}
    \nabla_{\phi} J_{MLE} = -2\nabla_{\phi}f(X;\bm{\phi})^{\top}(\bm{y}-f(X;\bm{\phi})) = 0.
\end{equation}
If the sensitivity matrix 
\begin{equation}\label{eq:sensitivity-matrix}
    \nabla_{\phi}f(X;\bm{\phi})\vert_{\phi}  = 
    \begin{bmatrix}
    \frac{\partial f}{\partial \phi_1}(X) & ... & \frac{\partial f}{\partial \phi_m}(X)
    \end{bmatrix}\vert_{\phi}
\end{equation}
has full column rank for all $\phi$, we may say that the problem is identifiable and there exist a unique solution to \eqref{eq:MLE}. Nevertheless, determining the rank of \eqref{eq:sensitivity-matrix} for the general nonlinear problem is nontrivial.

The sensitivity matrix is highly influenced by the span, or variability, of $X$, and uncertainty in the measurements of the explanatory variables. If the model is non-identifiable, a natural approach is to increase the variability of $X$. However, for petroleum production systems this is not trivial as the production data is influenced by the operational practices of the operator. Further, designed experiments are often unaffordable due to the operation of the asset at a non-optimal operating point. Other approaches to obtain identifiability of non-identifiable over-parameterized models are parameter ranking methods \citep{Chan1992} and parameter reduction methods \citep{FluidMech}. The first method estimates the parameters that are most influential to the model output while fixing the remaining parameters at constant prior values. The latter method removes redundant parameters. Nevertheless, in a real-life setting, the underlying process and parameter values are unknown, and fixing parameters at constant values may lead to bias estimates. For hybrid models, many of the parameters are non-physical. Therefore, finding redundant parameters or good priors to fix the less influential parameters is challenging. 

On the other hand, even if the sensitivity matrix has full rank, identifiability may be influenced by other aspects. For instance, for highly nonlinear, non-convex problems, iterative, stochastic, and local optimization algorithms are often required to solve the estimation problem in reasonable time \citep{Bottou2018}. The stochasticity and locality of these algorithms may prevent model identifiability. For instance, with stochastic gradient descent (SGD) the parameters are updated with
\begin{equation}\label{eq:SGD}
\bm{\phi}^{k+1} = \bm{\phi}^k - \alpha^k \mathcal{M}^k(\bm{y}^k, \bm{\hat{y}}^k),
\end{equation}
where $\alpha\leq1$ is the learning rate and $\mathcal{M}$ is a stochastic gradient commonly calculated using a subset or mini-batch of the data samples. When a model increases in complexity, both in terms of model parameters and available process data, batch SGD methods scale better and are computationally feasible \citep{Bottou2018}. The totality of these issues implies that model identifiability is challenging to obtain in practice for complex models, and the physical interpretability of the model may be easily lost. 

An alternative approach to counteract the loss of physical interpretability in non-identifiable models is using parameter regularization in the estimation problem. Regularization of the parameters is achieved by setting up maximum a posterior (MAP) estimation of the parameters, instead of MLE. MAP estimation attempts to find the mode of the posterior probability distribution of the model parameters given the data $\mathcal{D}$
\begin{equation}
    \bm{\phi}^*_{MAP} =\arg \max_{\bm{\phi}} p(\bm{\phi} | \mathcal{D}).
\end{equation}
Through utilization of Bayes' theorem and assuming normally distributed parameter priors $\phi_i \sim \mathcal{N}(\mu_i, \sigma_i^2), i\in\{1,..,m\}$, the optimization problem
\begin{equation}\label{eq:MAP}
\begin{array}{ll}
 &\bm{\phi}^*_{MAP} = \arg \min_{\bm{\phi}}\Bigg[\underbrace{\left(\bm{y} - f(X; \bm{\phi})\right)^{\top}\left(\bm{y} - f(X; \bm{\phi})\right)}_{\text{MLE}} \\
 &+ \underbrace{ (\bm{\phi}- \bm{\mu})^{\top} \Pi (\bm{\phi}- \bm{\mu})}_{\text{parameter regularization}}\Bigg], \quad\Pi = diag(\frac{\sigma_{\epsilon}^2}{\sigma_1^2},..,\frac{\sigma_{\epsilon}^2}{\sigma_{m}^2})
\end{array}
\end{equation}
is derived. In short, MAP estimation is a trade-off between minimizing the deviation between model estimates and measurements, and penalization of the deviation of the parameters away from their prior mean value $\bm{\mu}$. This type of regularization method is called Tikhonov or $\ell_2$-regularization and is referred to as a shrinking method \citep{Hastie2009}. Dependent on $\Pi$, $\ell_2$-regularization may make the MAP problem nonsingular with one unique solution when the MLE problem is not. To see this, consider the linear case where
\begin{equation}\label{eq:f-lin}
    \bm{\hat{y}} = f(X; \bm{\phi}) = X\bm{\phi}.
\end{equation}
The solutions of MLE in \eqref{eq:MLE}, and MAP in \eqref{eq:MAP}, for \eqref{eq:f-lin} are respectively
\begin{equation}\label{eq:map-linear}
\begin{array}{ll}
    \bm{\phi}^{\star}_{MLE} = \left(X^{\top}X\right)^{-1}X^{\top}\bm{y},\\ 
    \bm{\phi}^{\star}_{MAP} = \left(\Pi + X^{\top}X\right)^{-1}\left(X^{\top}\bm{y} + \Pi\bm{\mu}\right).
\end{array}
\end{equation}
The MAP solution adds a diagonal, positive definite matrix to $X^{\top}X$ before inversion. Dependent on $\Pi$, the MAP problem may be nonsingular even if the $X^{\top}X$ does not have full column rank. If the elements in $\Pi$ are set sufficiently high, the MAP solution approximate to 
\begin{equation}
    \bm{\phi}^{\star}_{MAP} \approx \bm{\mu}.
\end{equation}
This indicates the importance of good parameter priors. An appropriate selection of $\Pi$ will allow parameter deviation away from the prior mean value while keeping the parameters within a feasible range, thereby preserving physical interpretability. The same results may be obtained (locally) for the general nonlinear system through a first-order Taylor approximation of $f$.

In this study, MAP estimation is used to train the mechanistic and hybrid models. For hybrid models, the parameter regularization term in \eqref{eq:MAP} is divided into two, one each for the physical and non-physical parameters. For the physical parameters, the prior parameter distributions $\{\mu_i, \sigma_i^2\}_{i=1}^m$ need to be specified. The variance of the parameters may be determined using physical bounds. Notice, if $\sigma_i^2\to\infty$ then $\Pi\approx\bm{0}$, and the MAP estimation becomes an MLE problem in practice. The same effect may be achieved by setting $\sigma_{\epsilon}^2$ small. Ideally, $\sigma_{\epsilon}^2$ should be determined prior to training, however, it may be used as a regularization tuning constant. For the non-physical parameters, common practice is to penalize large parameters by setting the same prior $\mu_i = 0$ and $\Pi = \lambda \mathbb{I}$ on all parameters \citep{Goodfellow2016}. The optimizer Adam \citep{adam} is utilized for all models. A hyperparameter search with Bayesian Optimization \citep{Klein2017} is used to find $\lambda$ and the learning rate $\alpha$ in \eqref{eq:SGD}. 

\section{Choke models}\label{sec:choke-models}
\begin{figure}
\centering
\includegraphics[width=4.5cm]{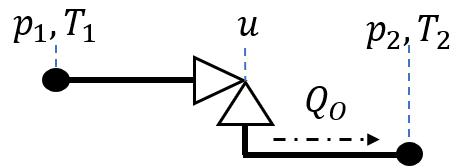}   
\caption{The petroleum well production choke valve with available measurement sensors.} 
\label{fig:choke}
\end{figure}

A petroleum well production choke valve may be illustrated as in Figure \ref{fig:choke}. Available sensor measurements are typically pressures ($p$) and temperatures ($T$), upstream (1) and downstream (2) the choke valve, and choke openings ($u$). In parameter estimation, measurements of the model output, in this research the oil volumetric flow rate, ($Q_O$), is also required, for instance from a multiphase flow meter or test separator. The mass fractions are treated as known and as inputs to the model. 
The mechanistic model used in this study is taken from \citet{Sachdeva1986}. The choke model is developed from the combined steady-state mass and momentum balance along a streamline and the equation for the mass flow rate through the choke is given as  
\begin{equation}\label{eq:mass_flow_rate}
\begin{array}{ll}
    \dot{m} = C_DA_2(u)\Bigg(&2\rho_2^2p_1\bigg(\frac{\kappa}{\kappa -1}\eta_G\left(\frac{1}{\rho_{G,1}}-\frac{p_r}{\rho_{G,2}}\right) \\
    &+ \left(\frac{\eta_O}{\rho_O} + \frac{\eta_W}{\rho_W}\right)(1 - p_r)\bigg)\Bigg)^{1/2},
\end{array}
\end{equation}
where $A$ is the flow area of the choke, typically a nonlinear function of $u$, $\rho$ is the mixture density, $\eta_i, \rho_i, i\in \{G, O, W\}$ are the mass fractions and densities of gas, oil, and water, respectively, $\kappa$ is the adiabatic gas expansion coefficient, $p_r$ is the downstream to upstream pressure ratio, and $C_D$ is a discharge coefficient commonly introduced to account for modeling errors. The oil volumetric flow rate in standard conditions is obtained with 
\begin{equation}\label{eq:volumetric-flow-rate-phase}
    Q_{O} = \frac{\eta_O\dot{m}}{\rho_{O,ST}}.
\end{equation}

In model development, \citet{Sachdeva1986} assumes 1) the upstream gas density may be described with the real gas law
\begin{equation}\label{eq:real-gas-law}
    \rho_{G1} = \frac{p_1M_G}{Z_1RT_1},
\end{equation}
where $M_G$ is the molar mass of gas, $Z$ is the gas compressibility factor, and $R$ is the universal gas constant, 2) the gas expansion across the choke may be assumed adiabatic
\begin{equation}\label{eq:polytropic-gas-expansion}
    \frac{1}{\rho_{G,2}} = \frac{1}{\rho_{G,1}}\left(\frac{p_1}{p_2}\right)^{\frac{1}{\kappa}},
\end{equation} 
3) the liquid is incompressible such that the oil and water densities remain constant across the choke, and 4) the mixture density may be assumed homogeneous
\begin{equation}\label{eq:homogeneous-mix-density}
    \frac{1}{\rho} = \frac{\eta_G}{\rho_G} + \frac{\eta_O}{\rho_O} + \frac{\eta_W}{\rho_W}, \quad \eta_W = 1-\eta_G -\eta_O.
\end{equation}
The \citeauthor{Sachdeva1986} model distinguishes between sub-critical and critical flow\footnote{Critical flow through a choke occur when a reduction in $p_2$ for a fixed $p_1$ does not increase the flow rate \citep{NodalAnalysis2015}.} through the choke restriction by defining the pressure ratio as
\begin{equation}\label{eq:pressure-ratio}
    p_r = 
    \begin{cases}
    \frac{p_2}{p_1} & \frac{p_2}{p_1} \geq p_{r,c} \\
    p_{r,c} & otherwise
    \end{cases}
\end{equation}
For multiphase flow, a rule of thumb for the critical pressure ratio is $p_{r,c} \approx 0.6$~\citep{NodalAnalysis2015}.

In this article, the \citeauthor{Sachdeva1986} choke model will be referred to as the mechanistic model (MM) and we specify the model parameters as
\begin{equation}\label{eq:theta-mm}
    \bm{\phi}_{MM} = [\rho_O, \rho_W, \kappa, M_G, p_{r,c}, C_D].
\end{equation}

To hybridize the MM, we introduce a neural network (NN) into the model equations. The NN is a collection of \textit{layers}, where each layer, $i\in\{1,..K\}$, outputs ($z_i$) a transformation of the inputs ($z_{i-1}$). We use a piece-wise linear transformation with weights $\bm{W}_i$ and bias $b_i$, and the rectified linear unit (ReLU) as activation function, $a$. 
\begin{equation}
\begin{aligned}
    \bm{z}_i &= a(\bm{W}_i\bm{z}_{i-1} + b_i)\\
    & = \max\{0, \bm{W}_i\bm{z}_{i-1} + b_i\}, \quad i\in\{1,...K\}
\end{aligned}
\end{equation}
In this research, the flow area function of the outlet, $A_2(u)$, will be represented with a neural network
\begin{equation}
    A_2 = g(u; \bm{\phi}_{DD}),
\end{equation}
where the choke opening is used as input to the network and the network parameters are the collection of weights and biases on all layers $\bm{\phi}_{DD} = \{(\bm{W}_1, b_1),...,(\bm{W}_K, b_K)\}$. The rest of the mechanistic equations from the MM remain as before. One may think of this model design as a way to alleviate the MM assumption of the shape of the area function. Naturally, relaxation of other MM assumptions such as the real gas law or adiabatic gas expansion is another possible model design. We incorporate the constant discharge coefficient $C_D$ into the above function such that the remaining physical parameters of the HM are
\begin{equation}\label{eq:theta-hm}
    \bm{\phi}_{HM} = [\rho_O, \rho_W, \kappa, M_G, p_{r,c}].
\end{equation}

\section{Case study - synthetic data}\label{sec:synthetic-cs}
In this case study, we investigate the identifiability and interpretability of the MM and HM presented in Section \ref{sec:choke-models}. To have perfect information about the underlying process, we generate noise-free, synthetic data of the oil volumetric flow rate through the choke with the MM from Section \ref{sec:choke-models}. A set of realistic combinations of the inputs $\bm{x}$ taken from two anonymous petroleum fields are used. If the model is identifiable, we should expect the physical parameters to converge to the true underlying parameter values of the data-generating process \emph{without} regularizing the parameters.  

Table \ref{tb:prior} gives the true parameter values of the data generating process and the prior parameter distributions used for initialization and regularization of the physical model parameters. The neural network depth and width in the HM are chosen $3\times100$, and the non-physical parameters are initialized with He-initialization \citep{He2015}. Due to the stochastic nature of the optimization algorithm, the models are trained several times, and the median mean absolute error (MAE), absolute percentage error (MAPE), and parameter values are reported. The errors are calculated using an independent test set.  
\begin{table}[h]
\centering
\caption{Values of the true parameters and the prior mean and standard deviation.}
\begin{tabular}{llll}
$\phi$  & True  & $\mu$ & $\sigma$\\ \hline      
$\rho_O$    & 760 & 800 & 33.3 \\
$\rho_W$  & 1010  & 1025 & 8.33 \\
$\kappa$  & 1.30  & 1.32 & 0.033 \\
$M_G$   & 0.021 & 0.027 & 0.003 \\
$p_{r,c}$ & 0.55 & 0.6 & 0.067 \\
$C_D$ & 1.0 & 0.9 & 0.25 \\\hline
\end{tabular}
\label{tb:prior}
\end{table}

\begin{figure}[ht]
\centering
\includegraphics[width=8cm]{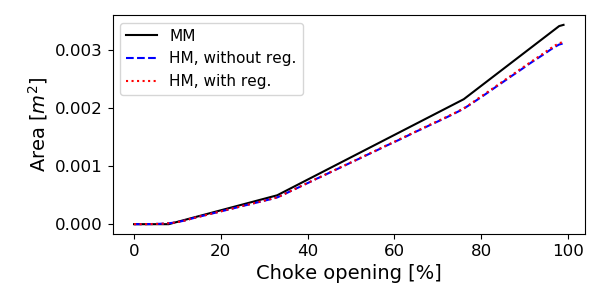}   
\caption{The neural network area function in the HM for different degree of regularization. The mechanistic area function is illustrated for reference.} 
\label{fig:area-function-synthetic}
\end{figure}

Initially, we set $\sigma_{\epsilon}$ small to obtain $\Pi \approx \bm{0}$, resulting in negligible regularization. It turns out that neither the MM nor the HM obtains convergence of the physical parameters to the true underlying values, see the two first columns of Table \ref{tb:results-synthetic}. Furthermore, some of the parameters converge to physically infeasible values. On the other side, through experimentation, it was found that fixing the value of $C_D$ in the MM at its true value resulted in the convergence of the remaining parameters to their true value. Nevertheless, in a real-life setting, the true underlying parameter values are unknown and the process will never be perfectly represented by a model. Further, the available data is often limited. Therefore, in particular for hybrid models with a large number of non-physical parameters, a better approach may be to include all uncertain parameters in the estimation problem instead of fixing some of them and use parameter regularization to ensure the preservation of the physical interpretability of the model. The two last columns of Table \ref{tb:results-synthetic} illustrates the results of the MM and HM with sufficient regularization of the model parameters. The learned HM area function with and without regularization is illustrated in Fig.~\ref{fig:area-function-synthetic}. Observe how the curve is close to that of the MM, thereby retaining the physical interpretability of the model. Pay in mind, it was not expected that the HM should obtain a lower error than the MM as the underlying model structure is perfectly known and matches the MM. In a practical case, one should expect an overall decrease in error due to the increased model flexibility of the HM. 

\begin{table}[ht]
\centering
\caption{The median error and parameter values of the MM and HM without regularization for the synthetic data.}
\begin{tabular}{lllll}
& \multicolumn{2}{c}{Without reg.} &\multicolumn{2}{c}{With reg.} \\
& MM & HM & MM & HM \\ \toprule
MAE &  0.0 & 0.2 & 0.0 & 0.1 \\ 
MAPE & 0.0 & 0.4 & 0.0 & 0.1 \\ \midrule
$\rho_O$  & 979 & 737 & 777 & 772\\
$\rho_W$  & 1302 & 976  & 1027 & 1025    \\
$\kappa$  & 1.31 & 1.30 & 1.30 & 1.30        \\
$M_G$     & 0.028 & 0.021 & 0.022 & 0.022 \\
$p_{r,c}$ & 0.55 & 0.55 & 0.55 & 0.55  \\ 
$C_D$     & 0.88 & \text{n.a.} & 0.99 & \text{n.a.}            \\\bottomrule
\end{tabular}
\label{tb:results-synthetic}
\end{table}

\section{Case study - Edvard Grieg}\label{sec:EG-cs}
In this case study, the physical interpretability and predictive performance are investigated for petroleum production choke valve models developed using real production data. We use historical data from six petroleum wells on the asset Edvard Grieg \citep{EG} and train the MM and the HM for each well. The data have been preprocessed in two steps. Firstly, the processing technology in \citep{Grimstad2016} is used to produce a compressed data set of steady-state operating points suitable for steady-state modeling. Secondly, a set of filters are applied to remove significant outliers such as negative pressures or flow rates. For each well, the historical data is split into training and test data, using the three latest months as test data. This is to mimic an industrial case where the model is used to predict future flow rates. 

\begin{figure}[ht]
\centering
\includegraphics[width=8cm]{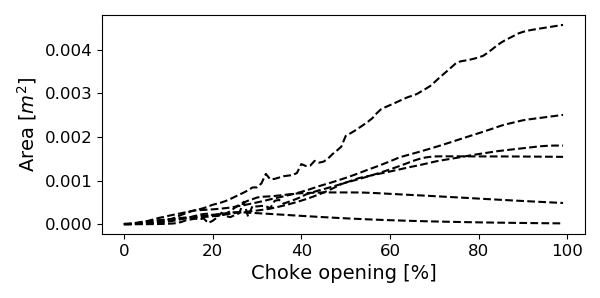}   
\caption{The neural network area function in the HM for the different wells. We see that there are large individual differences for each well.} 
\label{fig:area-function-eg}
\end{figure}

As in Section \ref{sec:synthetic-cs}, the physical parameter initial values are drawn from prior parameter normal distributions, see Table \ref{tb:prior}, and regularization is applied to enforce convergence of the physical parameters within feasible bounds. As before, the models are trained several times due to the stochasticity of the estimation problem. Table \ref{tb:results-eg} gives the minimum, median, and maximum MAE, MAPE, and parameter values across the six wells grouped on model type. Fig.~\ref{fig:area-function-eg} illustrates the learned neural network area function in the HM for the six wells. 
\begin{table}[h]
\centering
\caption{The minimum, median, and maximum error and parameter values for the six wells, grouped on model type.}
\begin{tabular}{lllllllll}
& \multicolumn{3}{c}{MM} & \multicolumn{3}{c}{HM} \\ \toprule
& min & med. & max & min & med. & max \\  \midrule
MAE & 3.4 & 8.8 & 19.9 & 2.3 & 4.8 & 11.5  \\ 
MAPE & 7.3 & 14.0 & 68.3 & 2.3 & 9.1 & 47.8 \\ \midrule
$\rho_O$  & 608 & 719 & 812 & 634& 682 & 707\\
$\rho_W$  & 1025 & 1025 & 1025 & 1025 & 1025 & 1025 \\
$\kappa$  & 1.29 & 1.33 & 1.49 & 1.33 & 1.35 & 1.36  \\
$M_G$     & 0.027 & 0.033 & 0.041 & 0.029 & 0.037 & 0.045\\
$p_{r,c}$ & 0.48 & 0.66 & 0.93 & 0.65 & 0.68 & 0.88  \\ 
$C_D$  & 0.72 &  0.84 & 0.91 & n.a. & n.a. & n.a. \\ \bottomrule
\end{tabular}
\label{tb:results-eg}
\end{table}

First of all, notice that the results indicate a significant reduction in prediction error comparing the HM to the MM. Also notice that the values for the physical parameters mostly stay feasible, with some exceptions. On the other hand, there are large variations in the errors for each individual model. The overall best performing model is an HM with 2.3\% MAPE whereas the overall worst performing model is an MM with 68.3\% MAPE. Furthermore, observe from Fig.~\ref{fig:area-function-eg} that some of the learned area functions are in line with the expected physical trend (an increase in the opening gives an increase in area) whereas other converges to zero. 

There may be multiple causes for the large variations. Firstly, if the available data are noisy, faulty, or lacking, the model performance and physical interpretability may be influenced. Investigations of the dataset show that for the wells where the learned neural network area function converges to zero, there are lacking measurements of the choke opening in certain operating regions. Pretraining the neural network could be beneficial in such situations. Further, inappropriate data may easily lead to overfitting of the model parameters and commonly results in poor predictive performance on unseen data. 

Secondly, the model structure may be inappropriate. The baseline mechanistic model is a simplified model and may not capture physical effects equally well in all flow regimes and operating conditions. Mechanistic models with higher complexity or models that include effects of the remaining production system such as the wellbore, would likely decrease the error. Further, in this research, the HM only alleviates the mechanistic model assumption of the area function shape. Other assumptions that could be alleviated are for instance the real gas law, the homogeneous mixture density, or the adiabatic gas expansion. Yet, introducing additional data-driven elements could influence model interpretability and should be carefully considered.

Lastly, the utilized process data originate from a non-stationary process, the reservoir. Therefore, using three months as test data in which the model and parameters remain constant may be inappropriate and may cause large prediction errors. Updating the model more frequently, for instance in an online learning fashion, would likely improve the performance.

\section{Conclusion}
This study argues that model identifiability is challenging to obtain in practice due to both deterministic and stochastic uncertainties. In particular in situations where the underlying process is complex and highly nonlinear, as is the case in petroleum production systems, and where the available data is limited. Even though identifiability is often thought of as a prerequisite for obtaining physically interpretable models, we have illustrated through an industrial case study that physical interpretability may be preserved for non-identifiable hybrid models by the inclusion of sufficient regularization of the model parameters in the estimation problem. Further, hybrid models may improve the predictive performance compared to a mechanistic model due to the increased model flexibility. This is demonstrated in a case study on real production data from six petroleum wells on the asset Edvard Grieg. The hybrid model decreases the median MAPE across the six wells by $35\%$ compared to a mechanistic model while staying physically interpretable. On the other hand, the study shows that the predictive performance and the physical interpretability are influenced by the available data, and there are large variations in the results on a well-level. Future research should look into other hybrid model variants and online learning to improve predictive performance and physical interpretability. Certainly, any general conclusions cannot be drawn as we have only experimented with two different model types and with historical data from one petroleum field. Experimentation with data from other wells and fields would greatly benefit the results in this study. 

\begin{ack}
This research is a part of BRU21 - NTNU Research and Innovation Program on Digital and Automation Solutions for the Oil and Gas Industry (\url{www.ntnu.edu/bru21}) and supported by Lundin Energy Norway.    
\end{ack}

\bibliography{ms}     

\end{document}